\documentclass[fleqn,10pt,mtlplain]{wlscirep}

\usepackage[utf8]{inputenc}
\usepackage[T1]{fontenc}
\usepackage{mathpazo}
\usepackage{inconsolata}
\usepackage{microtype}

\makeatletter
\renewcommand\AB@authnote[1]{\textsuperscript{\sffamily\mdseries #1}}
\renewcommand\AB@affilnote[1]{\textsuperscript{\sffamily\mdseries #1}}
\makeatother

\usepackage{bm}
\usepackage{xspace}
\usepackage[normalem]{ulem}
\usepackage{gensymb}
\usepackage[symbol]{footmisc}
\usepackage{siunitx}
\usepackage{doi}

\usepackage{cleveref}
\crefname{figure}{Figure}{Figures}
\crefname{table}{Table}{Tables}
\crefname{chapter}{Chapter}{Chapters}
\crefname{section}{Section}{Sections}

\definecolor{lightmagenta}{RGB}{250,150,200}

\newcommand{\recommendation}[1]{\vspace{\baselineskip}\noindent\emph{Recommendation:} #1}
\newcommand{\pvalue}{\emph{p}-value\xspace}

\newcommand{\aachen}{Institute for Theoretical Particle Physics and Cosmology (TTK), RWTH Aachen University, Sommerfeldstra\ss e 14, D-52056 Aachen, Germany}
\newcommand{\queens}{Department of Physics, Engineering Physics and Astronomy, Queen's University, Kingston ON K7L 3N6, Canada}
\newcommand{\imperial}{Department of Physics, Imperial College London, Blackett Laboratory, Prince Consort Road, London SW7 2AZ, UK}
\newcommand{\cambridge}{Cavendish Laboratory, University of Cambridge, JJ Thomson Avenue, Cambridge, CB3 0HE, UK}
\newcommand{\oslo}{Department of Physics, University of Oslo, Box 1048, Blindern, N-0316 Oslo, Norway}
\newcommand{\adelaide}{ARC Centre for Dark Matter Particle Physics, Department of Physics, University of Adelaide, Adelaide, SA 5005, Australia}
\newcommand{\louvain}{Centre for Cosmology, Particle Physics and Phenomenology (CP3), Universit\'{e} catholique de Louvain, B-1348 Louvain-la-Neuve, Belgium}
\newcommand{\monash}{School of Physics and Astronomy, Monash University, Melbourne, VIC 3800, Australia}
\newcommand{\mcdonald}{Arthur B. McDonald Canadian Astroparticle Physics Research Institute, Kingston ON K7L 3N6, Canada}
\newcommand{\nanjing}{Department of Physics and Institute of Theoretical Physics, Nanjing Normal University, Nanjing, Jiangsu 210023, China}
\newcommand{\okc}{Oskar Klein Centre for Cosmoparticle Physics, AlbaNova University Centre, SE-10691 Stockholm, Sweden}
\newcommand{\perimeter}{Perimeter Institute for Theoretical Physics, Waterloo ON N2L 2Y5, Canada}
\newcommand{\uq}{School of Mathematics and Physics, The University of Queensland, St.\ Lucia, Brisbane, QLD 4072, Australia}
\newcommand{\gottingen}{Institut f\"ur Astrophysik und Geophysik, Georg-August-Universit\"at G\"ottingen, Friedrich-Hund-Platz~1, D-37077 G\"ottingen, Germany}

\newcommand{\kicc}{Kavli Institute for Cosmology, University of Cambridge, Madingley Road, Cambridge, CB3 0HA, UK}

\newcommand{\bonn}{University of Bonn, Physikalisches Institut, Nussallee 12, D-53115 Bonn, Germany}
\newcommand{\bom}{Bureau of Meteorology, Melbourne, VIC 3001, Australia}
\newcommand{\glasgow}{School of Physics and Astronomy, University of Glasgow, University Place, Glasgow, G12~8QQ, UK}
\newcommand{\sfu}{Department of Physics, Simon Fraser University, 8888 University Drive, Burnaby B.C., Canada}
\newcommand{\zzu}{School of Physics, Zhengzhou University, ZhengZhou 450001, China}
\newcommand{\lyon}{Universit\'e de Lyon, Universit\'e Claude Bernard Lyon 1, CNRS/IN2P3, Institut de Physique des 2 Infinis de Lyon, UMR 5822, F-69622, Villeurbanne, France}
\newcommand{\cernth}{Theoretical Physics Department, CERN, CH-1211 Geneva 23, Switzerland}
\newcommand{\cernex}{Experimental Physics Department, CERN, CH–1211 Geneva 23, Switzerland}
\newcommand{\infnt}{Istituto Nazionale di Fisica Nucleare, Sezione di Torino, via P. Giuria 1, I–10125 Torino, Italy}
\newcommand{\ifj}{Institute of Nuclear Physics, Polish Academy of Sciences, Krakow, Poland}
\newcommand{\ific}{Instituto de F\'isica Corpuscular, IFIC-UV/CSIC, Apt.\ Correus 22085, E-46071, Valencia, Spain}
\newcommand{\kansas}{Department of Physics and Astronomy, University of Kansas, Lawrence, KS 66045, USA}
\newcommand{\wsu}{Department of Physics, Weber State University, 1415 Edvalson St., Dept. 2508, Ogden, UT 84408, USA}
\newcommand{\tum}{Physik Department T70, James-Franck-Stra{\ss}e, Technische Universit\"at M\"unchen, D-85748 Garching, Germany}
\newcommand{\uppsala}{Department of Physics and Astronomy, Uppsala University, Box 516, SE-751 20 Uppsala, Sweden}
\newcommand{\desy}{Deutsches Elektronen-Synchrotron DESY, Notkestr.~85, 22607 Hamburg, Germany}
\newcommand{\kit}{Institut f\"ur Theoretische Teilchenphysik, Karlsruhe Institute of Technology, D-76131 Karlsruhe, Germany}
\newcommand{\ucla}{Physics and Astronomy Department, University of California, Los Angeles, CA 90095, USA}

\title{Simple and statistically sound recommendations for analysing physical theories}

\preprint{
PSI-PR-20-23;                    
BONN-TH-2020-11;                 
CP3-20-59;                       
KCL-PH-TH/2020-75;               
P3H-20-080; TTP20-044;           
TUM-HEP-1310/20;                 
IFT-UAM/CSIC-20-180;             
TTK-20-47;                       
CERN-TH-2020-215;                
FTPI-MINN-20-36; UMN-TH-4005/20; 
HU-EP-20/37;                     
DESY 20-222;                     
ADP-20-33/T1143;                 
Imperial/TP/2020/RT/04;          
UCI-TR-2020-19                   
}

\author[1]{Shehu~S.~AbdusSalam}
\author[a,2,3]{Fruzsina~J.~Agocs}
\author[4]{Benjamin~C.~Allanach}
\author[a,5,6]{Peter~Athron}
\author[a,6]{Csaba~Bal{\'a}zs}
\author[b,7]{Emanuele Bagnaschi}
\author[c,8]{Philip Bechtle}
\author[b,9]{Oliver Buchmueller}
\author[a,10]{Ankit~Beniwal}
\author[a,11]{Jihyun Bhom}
\author[a,9,12]{Sanjay Bloor}
\author[a,13]{Torsten Bringmann}
\author[a,14]{Andy~Buckley}
\author[15]{Anja~Butter}
\author[a,16]{Jos{\'e}~Eliel~Camargo-Molina}
\author[a,11]{Marcin Chrzaszcz}
\author[a,17]{Jan Conrad}
\author[a,18]{Jonathan~M.~Cornell}
\author[a,19]{Matthias~Danninger}
\author[d,20]{Jorge de Blas}
\author[b,21]{Albert De Roeck}
\author[c,8]{Klaus Desch}
\author[b,22]{Matthew Dolan}
\author[c,8]{Herbert Dreiner}
\author[d,23]{Otto Eberhardt}
\author[b,24]{John Ellis}
\author[a,9,25]{Ben~Farmer}
\author[d,26]{Marco~Fedele}
\author[b,27]{Henning Fl{\"a}cher}
\author[a,5,*]{Andrew~Fowlie}
\author[a,6]{Tom{\'a}s~E.~Gonzalo}
\author[a,28]{Philip~Grace}
\author[c,8]{Matthias Hamer}
\author[a,2,3]{Will~Handley}
\author[a,29]{Julia~Harz}
\author[b,30]{Sven Heinemeyer}
\author[a,31]{Sebastian~Hoof}
\author[a,9]{Selim~Hotinli}
\author[a,28]{Paul~Jackson}
\author[a,32]{Felix~Kahlhoefer}
\author[e,33]{Kamila Kowalska}
\author[c,32]{Michael Kr\"amer}
\author[a,13]{Anders~Kvellestad}
\author[b,34]{Miriam Lucio Martinez}
\author[a,35,36]{Farvah~Mahmoudi}
\author[b,37]{Diego Martinez Santos}
\author[a,38]{Gregory~D.~Martinez}
\author[d,39]{Satoshi Mishima}
\author[b,40]{Keith Olive}
\author[d,41,42]{Ayan Paul}
\author[a,8]{Markus~Tobias~Prim}
\author[c,43]{Werner Porod}
\author[a,13]{Are~Raklev}
\author[a,9,12,17]{Janina~J.~Renk}
\author[a,44]{Christopher~Rogan}
\author[e,45,33]{Leszek Roszkowski}
\author[a,30]{Roberto~Ruiz~de~Austri}
\author[b,46]{Kazuki Sakurai}
\author[a,47]{Andre Scaffidi}
\author[a,9,12]{Pat~Scott}
\author[e,33]{Enrico~Maria~Sessolo}
\author[c,41]{Tim Stefaniak}
\author[a,32]{Patrick~St{\"o}cker}
\author[a,28,48]{Wei~Su}
\author[e,45,33]{Sebastian Trojanowski}
\author[9,49]{Roberto~Trotta}
\author[50]{Yue-Lin Sming Tsai}
\author[a,13]{Jeriek~Van~den~Abeele}
\author[d,51]{Mauro Valli}
\author[a,52,53,54]{Aaron~C.~Vincent}
\author[b,41,55]{Georg~Weiglein}
\author[a,28]{Martin~White}
\author[c,8]{Peter Wienemann}
\author[a,5]{Lei~Wu}
\author[a,6,56]{Yang~Zhang}

\affil[a]{The GAMBIT Community}
\affil[b]{The MasterCode Collaboration}
\affil[c]{The Fittino Collaboration}
\affil[d]{HEPfit}
\affil[e]{BayesFits Group\newline}

\affil[1]{Department of Physics, Shahid Beheshti University, Tehran, Iran}
\affil[2]{\cambridge}
\affil[3]{\kicc}
\affil[4]{DAMTP, University of Cambridge, Cambridge, CB3 0WA, UK}
\affil[5]{\nanjing}
\affil[6]{\monash}
\affil[7]{Paul Scherrer Institut, CH-5232 Villigen, Switzerland}
\affil[8]{\bonn}
\affil[9]{\imperial}
\affil[10]{\louvain}
\affil[11]{\ifj}
\affil[12]{\uq}
\affil[13]{\oslo}
\affil[14]{\glasgow}
\affil[15]{Institut f\"ur Theoretische Physik, Universit\"at Heidelberg, Germany}
\affil[16]{\uppsala}
\affil[17]{\okc}
\affil[18]{\wsu}
\affil[19]{\sfu}
\affil[20]{Institute of Particle Physics Phenomenology, Durham University, Durham DH1 3LE, UK}
\affil[21]{\cernex}
\affil[22]{ARC Centre of Excellence for Dark Matter Particle Physics, School of Physics, The University of Melbourne, Victoria 3010, Australia}
\affil[23]{\ific}
\affil[24]{Theoretical Particle Physics and Cosmology Group, Department of Physics, King’s College London, London WC2R 2LS, UK}
\affil[25]{\bom}
\affil[26]{\kit}
\affil[27]{H.~H.~Wills Physics Laboratory, University of Bristol, Tyndall Avenue, Bristol BS8 1TL, UK}
\affil[28]{\adelaide}
\affil[29]{\tum}
\affil[30]{Instituto de F\'isica Te\'orica UAM-CSIC, Cantoblanco, 28049, Madrid, Spain}
\affil[31]{\gottingen}
\affil[32]{\aachen}
\affil[33]{National Centre for Nuclear Research, ul. Pasteura 7, PL-02-093 Warsaw, Poland}
\affil[34]{Nikhef National Institute for Subatomic Physics, Amsterdam, Netherlands}
\affil[35]{\lyon}
\affil[36]{\cernth}
\affil[37]{Instituto Galego de F{\'i}sica de Altas Enerx{\'i}as, Universidade de Santiago de Compostela, Spain}
\affil[38]{\ucla}
\affil[39]{Theory Center, IPNS, KEK, Tsukuba, Ibaraki 305-0801, Japan}
\affil[40]{William I. Fine Theoretical Physics Institute, School of Physics and Astronomy, University of Minnesota, Minneapolis, MN 55455, USA}
\affil[41]{\desy}
\affil[42]{Institut f\"ur Physik, Humboldt-Universit\"at zu Berlin, D-12489 Berlin, Germany}
\affil[43]{University of W\"urzburg, Emil-Hilb-Weg 22, D-97074 Würzburg, Germany}
\affil[44]{\kansas}
\affil[45]{Astrocent, Nicolaus Copernicus Astronomical Center Polish Academy of Sciences, Bartycka 18, PL-00-716 Warsaw, Poland}
\affil[46]{Institute of Theoretical Physics, Faculty of Physics, University of Warsaw, ul. Pasteura 5, PL-02-093 Warsaw, Poland}
\affil[47]{\infnt}
\affil[48]{Korea Institute for Advanced Study, Seoul 02455, Korea}
\affil[49]{SISSA International School for Advanced Studies, Via Bonomea 265, 34136, Trieste, Italy}
\affil[50]{Key Laboratory of Dark Matter and Space Astronomy, Purple Mountain Observatory, Chinese Academy of Sciences, Nanjing 210033, China}
\affil[51]{Department of Physics and Astronomy, University of California, Irvine, California 92697, USA}
\affil[52]{\queens}
\affil[53]{\mcdonald}
\affil[54]{\perimeter}
\affil[55]{Institut f\"ur Theoretische Physik, Universit\"at Hamburg,Luruper Chaussee 149, 22761 Hamburg, Germany}
\affil[56]{\zzu}

\affil[*]{E-mail: andrew.j.fowlie@njnu.edu.cn}

\begin{abstract}
  Physical theories that depend on many parameters or are tested against data from many different experiments pose unique challenges to statistical inference. Many models in particle physics, astrophysics and cosmology fall into one or both of these categories. These issues are often sidestepped with statistically unsound \textit{ad~hoc} methods, involving intersection of parameter intervals estimated by multiple experiments, and random or grid sampling of model parameters. Whilst these methods are easy to apply, they exhibit pathologies even in low-dimensional parameter spaces, and quickly become problematic to use and interpret in higher dimensions. In this article we give clear guidance for going beyond these procedures, suggesting where possible simple methods for performing statistically sound inference, and recommendations of readily-available software tools and standards that can assist in doing so. Our aim is to provide any physicists lacking comprehensive statistical training with recommendations for reaching correct scientific conclusions, with only a modest increase in analysis burden.  Our examples can be reproduced with the code publicly available at \href{https://doi.org/10.5281/zenodo.4322283}{Zenodo}.
\end{abstract}

\begin{document}

\flushbottom
\maketitle

\clearpage
\thispagestyle{empty}

\section{Introduction}

The search for new particles is underway in a wide range of high-energy, astrophysical and precision experiments. These searches are made harder by the fact that theories for physics beyond the Standard Model almost always contain unknown parameters that cannot be uniquely derived from the theory itself. For example, in particle physics models of dark matter, these would be the dark matter mass and its couplings. Models usually make a range of different experimental predictions depending on the assumed values of their unknown parameters.  Despite an ever-increasing wealth of experimental data, evidence for specific physics beyond the Standard Model has not yet emerged, leading to the proposal of increasingly complicated models. This increases the number of unknown parameters in the models, leading to high-dimensional parameter spaces. This problem is compounded by additional calibration and nuisance parameters that are required as experiments become more complicated. Unfortunately, high-dimensional parameter spaces, and the availability of relevant constraints from an increasing number of experiments, expose flaws in the simplistic methods sometimes employed in phenomenology to assess models. In this article, we recommend alternatives suitable for today's models and data, consistent with established statistical principles.

When assessing a model in light of data, physicists typically want answers to two questions: \textit{a})~Is the model favoured or allowed by the data? \textit{b})~What values of the unknown parameters are favoured or allowed by the data? In statistical language, these questions concern model testing and parameter estimation, respectively.
Parameter estimation allows us to understand what a model could predict, and design future experiments to test it. On the theory side, it allows us to construct theories that contain the model and naturally accommodate the observations. Model testing, on the other hand, allows us to test whether data indicate the presence of a new particle or new phenomena.

Many analyses of particle physics models suffer from two key deficiencies. First, they overlay exclusion curves from experiments and, second, they perform a random or grid scan of a high-dimensional parameter space. These techniques are often combined to perform a crude hypothesis test. In this article, we recapitulate relevant statistical principles, point out why both of these methods give unreliable results, and give concrete recommendations for what should be done instead. Despite the prevalence of these problems, we stress that there is diversity in the depth of statistical training in the physics community. Physicists contributed to major developments in statistical theory\cite{Jeffreys:1939xee,2008arXiv0808.2902R} and there are many statistically rigorous works in particle physics and related fields, including the famous Higgs discovery,\cite{Chatrchyan:2012ufa,Aad:2012tfa} and global fits of electroweak data.\cite{Baak:2014ora} Our goal is to make clear recommendations that would help lift all analyses closer to those standards, though we urge particular caution when testing hypotheses as unfortunately there are no simple recipes. The examples that we use to illustrate our recommendations can be reproduced with the code publicly available through \href{https://doi.org/10.5281/zenodo.4322283}{Zenodo}.\cite{zenodo_record}

Our discussion covers both Bayesian methods,\cite{giulio2003bayesian,gregory2005bayesian,sivia2006data,Trotta:2008qt,von2014bayesian,bailer2017practical} in which one directly considers the plausibility of a model and regions of its parameter space, and frequentist methods,\cite{lyons1989statistics,cowan1998statistical,james2006statistical,behnke2013data}
in which one compares the observed data to data that could have been observed in identical repeated experiments.%
\footnote{We cite here introductory textbooks about statistics by and for scientists. Refs.~\citen{sivia2006data,james2006statistical} are particularly concise.}
Our recommendations are agnostic about the relative merits of the two sets of methods, and apply whether one is an adherent of either form, or neither.
Both approaches usually involve the so-called likelihood function,\cite{Cousins:2020ntk} which tells us the probability of the observed data, assuming a particular model and a particular combination of numerical values for its unknown parameters.

In the following discussions, we assume that a likelihood is available and consider inferences based on it. In general, though, the likelihood alone is not enough in frequentist inference (as well as for reference priors and some methods in Bayesian statistics that use simulation). One requires the so-called sampling distribution; this is similar to the likelihood function, except that the data is not fixed to the observed data (see the likelihood principle\cite{berger1988likelihood} for further discussion). There are, furthermore, situations in which the likelihood is intractable. In such cases, likelihood-free techniques may be possible.\cite{Brehmer:2020cvb} In fact, in realistic applications in physics, the complete likelihood is almost always intractable. Typically, however, we create summaries of the data by e.g.\ binning collider events into histograms. 

\begin{figure}[t]
  \centering
  \includegraphics[width=12cm]{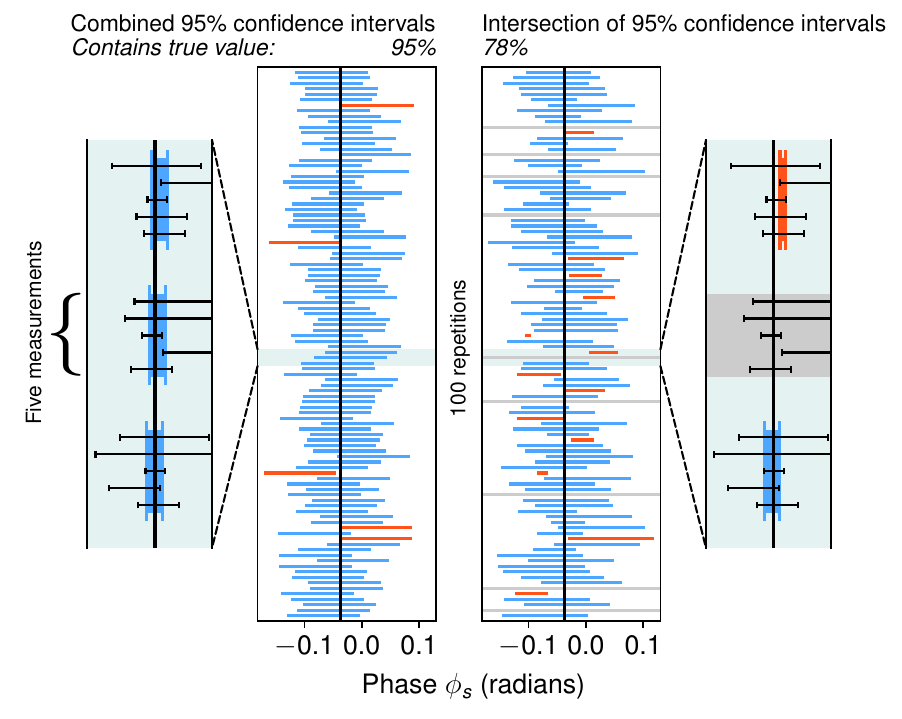}
  \caption{Confidence intervals in \num{100} pseudo-experiments, from the combination of five measurements~(\textit{left}) or from the intersection of five individual confidence intervals~(\textit{right}). We show the true value of~$\phi_s$ with a vertical black line. Intervals that contain the true value are shown in blue; those that do not are shown in red. On the right-hand side, grey bands indicate cases where no value can be found where the 95\% intervals from all five measurements overlap. Each bar originates from five pseudo-measurements, as shown zoomed-in to the side for a few points.}
  \label{fig:phi_s}
\end{figure}

\section{Problems of overlaying exclusion limits}\label{sec:limit_intersection}

Experimental searches for new phenomena are usually summarised by confidence regions, either for a particular model's parameters or for model-independent quantities more closely related to the experiment that can be interpreted in any model. For example, experiments performing direct searches for dark matter\cite{Undagoitia:2015gya} publish confidence regions for the mass and scattering cross section of the dark matter particle, rather than for any parameters included in the Lagrangian of a specific dark matter model. To apply those results to a given dark matter model, the confidence regions must be transformed to the parameter space of the specific model of interest. This can sometimes modify the statistical properties of the confidence regions, so care must be taken in performing the transformation.\cite{Bridges:2010de,Akrami11coverage,Strege12}

In the frequentist approach, if an experiment that measured a parameter were repeated over and over again, each repeat would lead to a different confidence region for the measured parameter. The coverage is the fraction of repeated experiments in which the resulting confidence region would contain the true parameter values.\cite{10.2307/91337}
The confidence level of a confidence region is the desired coverage.\footnote{Note that for discrete observations\cite{2010NIMPA.612..388C} or in the presence of nuisance parameters,\cite{Rolke:2004mj,Punzi:2005yq} confidence regions are often defined to include the true parameter values in \emph{at least} e.g.\ $95\%$ of repeated experiments,\cite{Zyla:2020zbs_conf_intervals} and that in some cases the nominal confidence level may not hold in practice.}
For example: a 95\% confidence region should contain the true values in 95\% of repeated experiments, and the rate at which we would wrongly exclude the true parameter values is controlled to be~5\%. Approximate confidence regions can often be found from the likelihood function alone using asymptotic assumptions about the sampling distribution, e.g., Wilks' theorem.\cite{wilks1938} However, it is important to check carefully that the required assumptions hold.\cite{Algeri:2019arh}

Confidence intervals may be constructed to be one- or two-tailed. By construction, in the absence of a new effect, a 95\% upper limit would exclude all effect sizes, including zero, at a rate of 5\%. The fact that confidence intervals may exclude effect sizes that the experiment had no power to discover was considered a problem in particle physics and lead to the creation of CL${}_s$ intervals.\cite{Read:2002hq} By construction, these intervals cannot exclude negligible effect sizes, and thus over-cover.

The analogous construct in Bayesian statistics is the credible region. First, prior information about the parameters and information from the observed data contained in the likelihood function are combined into the posterior using Bayes' theorem. Second,  parameters that are not of interest are integrated over, resulting in a marginal posterior distribution. A 95\% credible region for the remaining parameters of interest is found from the marginal posterior by defining a region containing 95\% of the posterior probability. In general, credible regions only guarantee average coverage: suppose we re-sampled model parameters and pseudo-data from the model and constructed 95\% credible regions. In 95\% of such trials, the credible region would contain the sampled model parameters.\cite{10.2307/2347266,james2006statistical} 
Whilst credible regions and confidence intervals are identical in some cases~(e.g.\ in normal linear  models), the fact that they in general lead to different inferences remains a point of contention.\cite{Morey2016} For both credible regions and confidence intervals, the level only stipulates the size of the region. One requires an ordering rule to decide which region of that size is selected. For example, the Feldman-Cousins construction\cite{Feldman:1997qc} for confidence regions and the highest-posterior density ordering rule for credible regions naturally switch from a one-\ to a two-tailed result.

When several experiments report confidence regions, requiring that the true value must lie within all of those regions amounts to approximating the combined confidence region by the intersection of regions from the individual experiments. This quickly loses accuracy as more experiments are applied in sequence, and leads to much greater than nominal error rates. This is because by taking an intersection of $n$ independent 95\% confidence regions, a parameter point has $n$ chances to be excluded at a $5\%$ error rate, giving an error rate of $1 - 0.95^n$.\cite{Junk:2020azi}

This issue is illustrated in \cref{fig:phi_s} using the $B$-physics observable $\phi_s$, which is a well-measured phase characterising CP-violation in $B_s$ meson decays.\cite{Amhis:2019ckw} We perform 10,000~pseudo-experiments.\footnote{In a pseudo-experiment, we simulate the random nature of a real experimental measurement using a pseudo-random number generator on a computer. Pseudo-experiments may be used to learn about the expected distributions of repeated measurements.} Each pseudo-experiment consists of a set of five independent Gaussian measurements of an assumed true Standard Model value of $\phi_s = -0.037$ with statistical errors
0.078, 
0.097, 
0.037, 
0.285, and 
0.17, 
which are taken from real ATLAS, CMS and LHCb measurements.\footnote{See Eq.~(91) and Table~22 in Ref.~\citen{Amhis:2019ckw}.}
We can then obtain the $95\%$ confidence interval from the combination of the five measurements in each experiment,\footnote{We used the standard weighted-mean approach to combine the results.\cite{Zyla:2020zbs_weighted_mean}} and compare it to the interval resulting from taking the intersection of the five $95\%$ confidence intervals from the individual measurements. We show the first \num{100} pseudo-experiments in \cref{fig:phi_s}. As expected, the $95\%$ confidence interval from the combination contains the true value in $95\%$ of simulated experiments. The intersection of five individual $95\%$ confidence intervals, on the other hand, contains the true value in only $78\%$ of simulations. Thus, overlaying regions leads to inflated error rates and can create a misleading impression about the viable parameter space. Whilst this is a one-dimensional illustration, an identical issue would arise for the intersection of higher-dimensional confidence regions. Clearly, rather than taking the intersection of reported results, one should combine likelihood functions from multiple experiments. Good examples can be found in the literature.\cite{Ciuchini:2000de, deAustri:2006jwj, Allanach:2007qk, Buchmueller:2011ab, Bechtle:2012zk, Fowlie:2012im, Athron:2017qdc}

\begin{figure}[t]
  \centering
  \includegraphics[width=14cm]{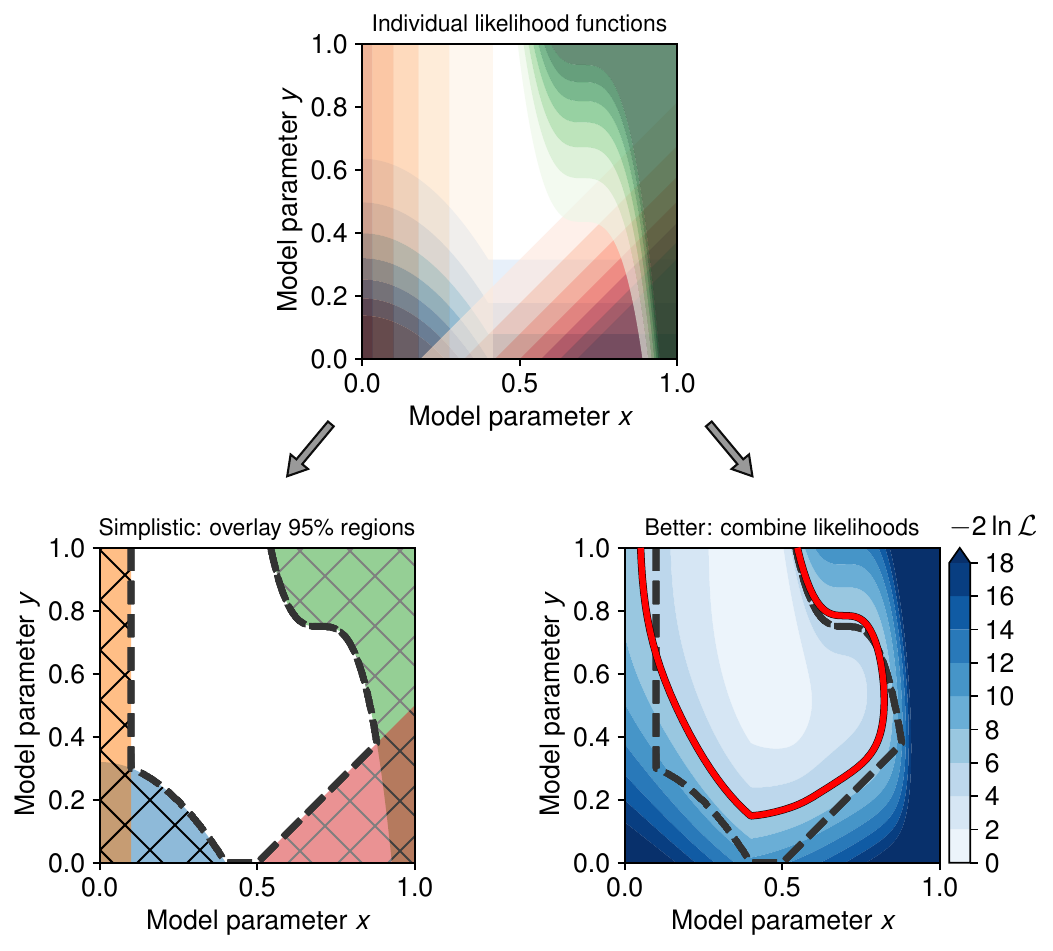}
  \caption{Starting from four individual likelihood functions~(\textit{top}; orange, blue, red and green, where lighter shades indicate greater likelihood), we compare overlaid $95\%$ contours~(\textit{bottom left}) versus a combination of the likelihoods~(\textit{bottom right}; blue contours). The dashed black line in both bottom panels is the intersection of the limits from the individual likelihoods. The red line in the bottom right panel is the resulting $95\%$ contour of the product of all likelihoods.}
  \label{fig:contours}
\end{figure}

In \cref{fig:contours} we again show the dangers of simply overlaying confidence regions. We construct several toy two-dimensional likelihood functions (top), and find their $95\%$ confidence regions (bottom left). In the bottom right panel, we show the contours of the combined likelihood function (blue) and a combined $95\%$ confidence region (red contour). We see that the intersection of confidence regions~(dashed black curve) can both exclude points that are allowed by the combined confidence region, and allow points that should be excluded. It is often useful to plot both the contours of the combined likelihood (bottom right panel) and the contours from the individual likelihoods (bottom left panel), in order to better understand how each measurement or constraint contributes to the final combined confidence region.

\recommendation{Rather than overlaying confidence regions, combine likelihood functions. Derive a likelihood function for all the experimental data (this may be as simple as multiplying likelihood functions from independent experiments), and use it to compute approximate joint confidence or credible regions in the native parameter space of the model.}

\section{Problems of uniform random sampling and grid scanning}\label{sec:random}

Parameter estimation generally involves integration of a posterior or maximisation of a likelihood function. This is required to go from the full high-dimensional model to the one or two dimensions of interest or to compare different models. In most cases this cannot be done analytically.  The likelihood function, furthermore, may be problematic in realistic settings. In particle physics,\cite{Balazs:2021uhg} it is usually moderately high-dimensional, and often contains distinct modes corresponding to different physical solutions, degeneracies in which several parameters can be adjusted simultaneously without impacting the fit, and plateaus in which the model is unphysical and the likelihood is zero.
On top of that, only noisy estimates of the likelihood may be available, such as from Monte Carlo simulations of collider searches for new particles, and derivatives of the likelihood function are usually unavailable.\cite{Balazs:2017moi} As even single evaluations of the likelihood function can be computationally expensive, the challenge is then to perform integration or maximisation in a high-dimensional parameter space using a tractable number of evaluations of the likelihood function.

\begin{figure}[t]
  \centering
  \includegraphics[width=15cm]{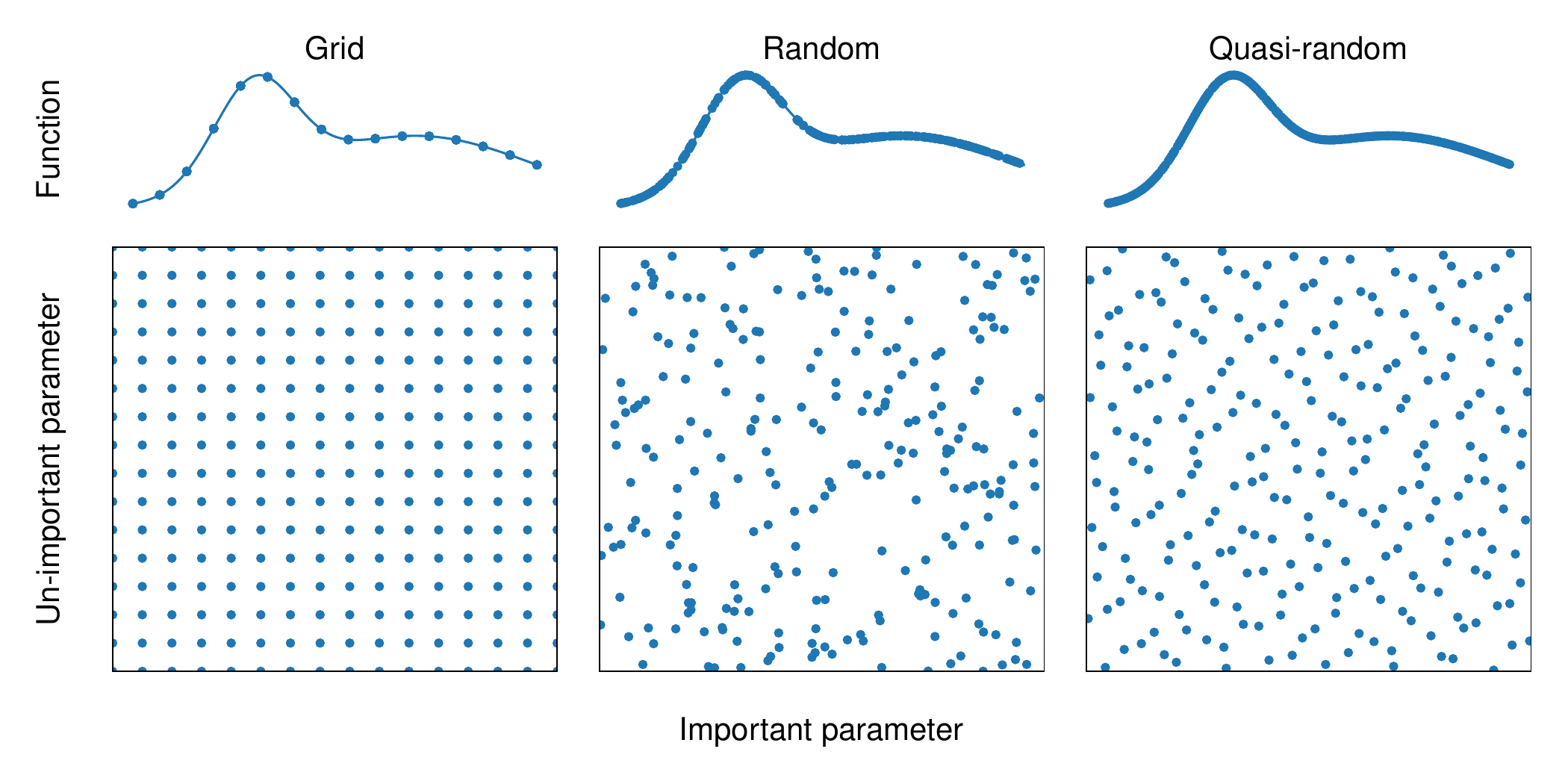}
  \caption{Grid, random and quasi-random sampling with 256 samples in two dimensions when the likelihood function is approximately one-dimensional. When the number of important parameters increases these methods perform poorly, as shown in \cref{fig:rosenbrock}.}
  \label{fig:quasi_random}
\end{figure}

Random and grid scans are common strategies in the high-energy phenomenology literature. In random scans, one evaluates the likelihood function at a number of randomly-chosen parameter points. Typically the parameters are drawn from a uniform distribution in each parameter in a particular parametrisation of the model, which introduces a dependency on the choice of parametrisation. In grid scans, one evaluates the likelihoods on a uniformly spaced grid with a fixed number of points per dimension.
It is then tempting to attribute statistical meaning to the number or density of samples found by random or grid scans. However, such an interpretation is very problematic, in particular when the scan is combined with the crude method described in \cref{sec:limit_intersection}, i.e.\ keeping only points that make predictions that lie within the confidence regions reported by every single experiment. 
It is worth noting that random scans often outperform grid scans: consider 100 likelihood evaluations in a two-parameter model where the likelihood function depends much more strongly on the first parameter than on the second. A random scan would try~100 different parameter values of the important parameter, whereas the grid scan would try just~10. In a similar vein, quasi-random samples that cover the space more evenly than truly random samples can out-perform truly random sampling.\cite{JMLR:v13:bergstra12a}
This is illustrated in \cref{fig:quasi_random} with 256 samples in two-dimensions.

However, random, quasi-random and grid scans are all extremely inefficient in cases with even a few parameters. The ``curse of dimensionality''~\cite{bellman1961adaptive} is one of the well-known problems: the number of samples required for a fixed resolution per dimension scales exponentially with dimension $D$: just~10 samples per dimension requires $10^D$ samples. This quickly becomes an impossible task in high-dimensional problems. Similarly, consider a $D$-dimensional model in which the interesting or best-fitting region occupies a fraction $\epsilon$ of each dimension. A random scan would find points in that region with an efficiency of $\epsilon^D$, i.e.\ random scans are exponentially inefficient. See Ref.~\citen{blum2020foundations} for further discussion and examples.

\begin{figure}[t]
  \centering
  \includegraphics[width=12cm]{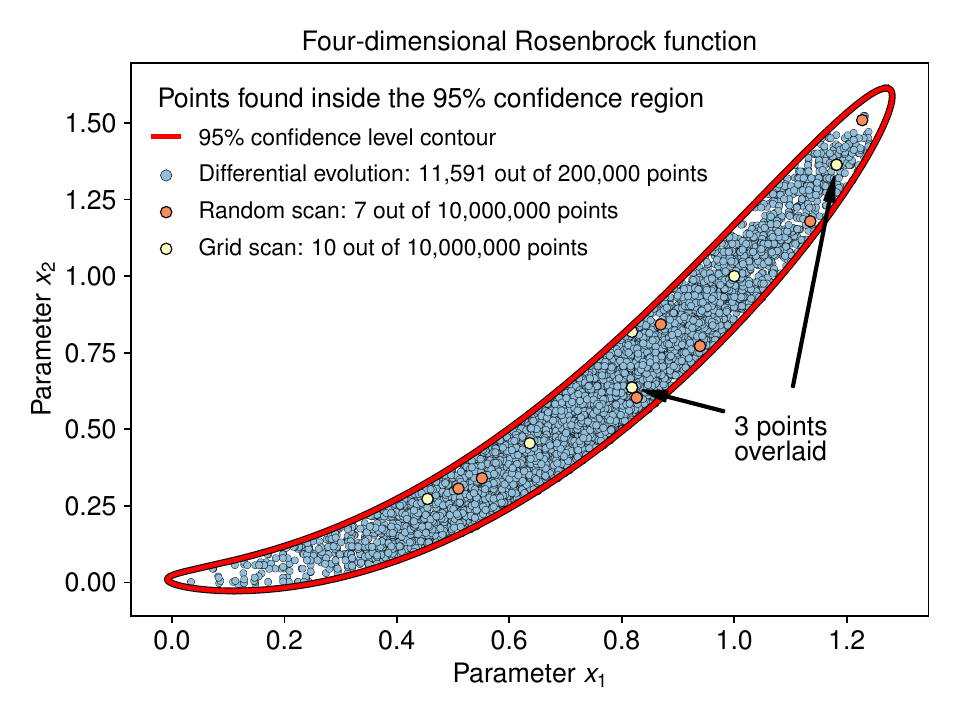}
  \caption{Points found inside the $95\%$ confidence region of the likelihood function, in a two-dimensional plane of the four-dimensional Rosenbrock problem. Points are shown from scans using differential evolution (blue), random sampling (orange) and grid sampling (yellow). For reference, we also show the actual $95\%$ confidence level contour of the likelihood function (red). Note that due to the projection of the four-dimensional space down to just two dimensions, two of the points shown from the grid sampler actually consist of three points each in the full four-dimensional space.}
  \label{fig:rosenbrock}
\end{figure}

These issues can be addressed by using more sophisticated algorithms that, for example, preferentially explore areas of the parameter space where the likelihood is larger. Which algorithm is best suited for a given study depends on the goal of the analysis. For Bayesian inference, it is common to draw samples from the posterior distribution or compute an integral over the model's parameter space, relevant for Bayesian model selection. See Ref.~\citen{2020arXiv200406425M} for a review of Bayesian computation. For frequentist inference, one might want to determine the global optimum and obtain samples from any regions in which the likelihood function was moderate. This can be more challenging than Bayesian computation. In particular, algorithms for Bayesian computation might not be appropriate optimizers. For example, Markov chain Monte Carlo methods draw from the posterior. In high-dimensions, the bulk of the posterior probability (the typical set) often lies well away from the maximum likelihood. This is another manifestation of the curse of dimensionality.

In \cref{fig:rosenbrock} we illustrate one such algorithm that overcomes the deficiencies of random and grid sampling and is suitable for frequentist inference. Here we assume that the logarithm of the likelihood function is given by a four-dimensional Rosenbrock function~\cite{10.1093/comjnl/3.3.175}
\begin{equation}\label{eq:rosenbrock}
  -2\ln\mathcal{L}(\bm{x}) = 2 \sum_{i=1}^3 f(x_i,\,x_{i+1}), \quad\text{where } f(a, b) = (1 - a)^2  + 100 \, (b - a^2)^2 \, .
\end{equation}
This is a challenging likelihood function with a global maximum at $x_i = 1$ ($i=1,2,3,4$).
We show samples found with $-2\ln\mathcal{L}(\bm{x}) \le 5.99$. This constraint corresponds to the two-dimensional $95\%$ confidence region, which in the $(x_1, x_2)$ plane has a banana-like shape~(red contour). We find the points using uniform random sampling from $-5$ to $5$ for each parameter (orange dots), using a grid scan (yellow dots), and using an implementation of the differential evolution algorithm~\cite{StornPrice95,2020SciPy-NMeth} operating inside the same limits (blue dots).
With only \num{2e5} likelihood calls, the differential evolution scan finds more than 11,500~points in the high-likelihood region,\footnote{We used a population size of 50 and stopped once the coefficient of variation of the fitness of the population dropped below 1\%. See the associated code for the complete settings.\cite{zenodo_record}}
whereas in \num{e7} tries the random scan finds only~7 high-likelihood samples, and the grid scan just~10. The random and grid scans would need over \num{e10} likelihood calls to obtain a similar number of high-likelihood points as obtained by differential evolution in just \num{2e5} evaluations. If likelihood calls are expensive and dominate the run-time, this could make differential evolution about \num{e5} times faster.

\recommendation{Use efficient algorithms to analyse parameter spaces, rather than grid or random scans. The choice of algorithm should depend on the goal. Good examples for Bayesian analyses are Markov chain Monte Carlo\cite{Hogg:2017akh, brooks2011handbook} and nested sampling.\cite{Skilling:2006gxv} Good examples for maximizing and exploring the likelihood are simulated annealing,\cite{Kirkpatrick671} differential evolution,\cite{StornPrice95} genetic algorithms\cite{1995ApJS..101..309C} and local optimizers such as Nelder-Mead.\cite{10.1093/comjnl/7.4.308} These are widely available in various public software packages.\cite{2020MNRAS.tmp..280S, Feroz:2008xx, Handley:2015fda, ForemanMackey:2012ig, Workgroup:2017htr, James:1975dr, hans_dembinski_2020_3951328, 2020SciPy-NMeth}}

\section{Problems with model testing}\label{sec:testing}

Overlaying confidence regions and performing random scans are straightforward
methods for ``hypothesis tests'' of physical theories with many parameters or testable predictions. For example, it is tempting to say that a model is excluded if a uniform random or grid scan finds no samples for which the experimental predictions lie inside every 95\%~confidence region. This procedure is, however, prone to misinterpretation: just as in \cref{sec:limit_intersection}, it severely under-estimates error rates, and, just as in \cref{sec:random}, it easily misses solutions.

Testing and comparing individual models in a statistically defensible manner is challenging and contentious. On the frequentist side, one can calculate a global \pvalue
: the probability of obtaining data as extreme or more extreme than observed, if the model in question is true. 
The \pvalue features in two distinct statistical approaches:\cite{doi:10.1198/0003130031856} first, the \pvalue may be interpreted as a measure of evidence against a model.\cite{fisher} See Refs.~\citen{Hubbard2008, doi:10.1080/00031305.1996.10474380,doi:10.1080/01621459.1987.10478397,Senn2001,Murtaugh2014} for discussion of this approach. Second, we may use the \pvalue to control the rate at which we would wrongly reject the model when it was true.\cite{10.2307/91247} If we reject when $p < \alpha$, we would wrongly reject at a rate $\alpha$. In particle physics, we adopt the $5\sigma$ threshold, corresponding to $\alpha \simeq \num{e-7}$.\cite{Lyons:2013yja}
When we compute \pvalue{}s, we should take into account all the tests that we might have performed. In the context of searches for new particles, this is known as the look-elsewhere effect. Whilst calculations can be greatly simplified by using asymptotic formulae,\cite{Cowan:2010js,Gross:2010qma} bear in mind that they may not apply.\cite{Algeri:2019arh} Also, care must be taken to avoid common misinterpretations of the \pvalue.\cite{GOODMAN2008135,Greenland2016} For example, the \pvalue is not the probability of the null hypothesis, or the probability that the observed data were produced by chance alone, or the probability of the observed data given the null hypothesis, or the rate at which we would wrongly reject the null hypothesis when it was true.

On the Bayesian side, one can perform Bayesian model comparison~\cite{Jeffreys:1939xee,Robert:1995oiy} to find any change brought about by data to the relative plausibility of two different models. The factor that updates the relative plausibility of two models is called a Bayes factor. The Bayes factor is a ratio of integrals that may be challenging to compute in high-dimensional models. Just as in Bayesian parameter inference, this requires constructing priors for the parameters of the two models, permitting one to coherently incorporate prior information. In this setting, however, inferences may be strongly prior dependent, even in cases with large data sets and where seemingly uninformative priors are used.\cite{berger2001objective,Cousins:2008gf} 
This sensitivity can be particularly problematic in high-dimensional models. Unfortunately, there is no unique notion of an uninformative prior representing a state of indifference about a parameter,\cite{Robert:1996lhi} though in special cases symmetry considerations may help.\cite{4082152}

Neither of these approaches is simple, either philosophically or computationally, and the task of model testing and comparison is in general full of subtleties. For example, they depend differently on the amount of data collected which leads to somewhat paradoxical differences between them.\!\!\cite{10.2307/2333251,Jeffreys:1939xee,Cousins:2013hry} See Refs.~\citen{Wagenmakers2007,doi:10.1177/1745691620958012,Benjamin2018,doi:10.1080/00031305.2018.1527253,Lakens2018} for recent discussions in other scientific settings. It is worth noting that there are connections between model testing and parameter inference in the case of nested models, i.e.\ when a model can be viewed as a subset of the parameter space of some larger, ``full'' model. A hypothesis test of a nested model can be equivalent to whether it lies inside a confidence region in the full model.\cite{kendall2a,Cousins:2018tiz} Similarly, the Bayes factor between nested models can be found from parameter inference in the full model alone through the Savage-Dickey ratio.\cite{10.2307/2958475} There are, furthermore, approaches beyond Bayesian model comparison and frequentist model testing that we do not discuss here.

\recommendation{In Bayesian analyses, carefully consider the choice of priors, their potential impact particularly in high-dimensions and check the prior sensitivity. In frequentist analyses, consider the look-elsewhere effect, check the validity of any asymptotic formulae and take care to avoid common misinterpretations of the \pvalue. If investigation of such subtleties fall outside the scope of the analysis, refrain from making strong statements on the overall validity of the theory under study.}

\section{Summary}\label{sec:conclude}

As first steps towards addressing the challenges posed by physical theories with many parameters and many testable predictions, we make three recommendations: \emph{i}) construct a composite likelihood that combines constraints from individual experiments, \emph{ii}) use adaptive sampling algorithms (ones that target the interesting regions) to efficiently sample the parameter spaces, and \emph{iii}) avoid strong statements on the viability of a theory unless a proper model test has been performed.
The second recommendation can be easily achieved through the use of any one of a multitude of publicly-available implementations of efficient sampling algorithms (for examples see \cref{sec:random}). For the first recommendation, composite likelihoods are often relatively simple to construct, and can be as straightforward as a product of Gaussians for multiple independent measurements. Even for cases where constructing the composite likelihood is more complicated, software implementations are often publicly available already.\cite{Athron:2017ard,hepfit,brinckmann2018montepython,Bhom:2020bfe,LikeDM,Collaboration:2242860,Aghanim:2019ame,IC79_SUSY,IC22Methods}

Given the central role of the likelihood function in analysing experimental data, it is in the interest of experimental collaborations to make their likelihood functions (or a reasonable approximation) publicly available to truly harness the full potential of their results when confronted with new theories. Even for large and complex datasets, e.g.~those from the Large Hadron Collider, there exist various recommended methods for achieving this goal.\cite{Cousins:451612,Vischia:2019uul,Abdallah:2020pec}

Our recommendations can be taken separately when only one of the challenges exists, or where addressing them all is impractical. However, when confronted with both high-dimensional models and a multitude of relevant experimental constraints, we recommend that they are used together to maximise the validity and efficiency of analyses.

\bibliography{stats,phys}

\section*{Acknowledgements}
BCA has been partially supported by the UK Science and Technology Facilities Council (STFC) Consolidated HEP theory grants ST/P000681/1 and ST/T000694/1. PA is supported by Australian Research Council (ARC) Future Fellowship FT160100274, and PS by FT190100814. PA, CB, TEG and MW are supported by ARC Discovery Project DP180102209. CB and YZ are supported by ARC Centre of Excellence CE110001104 (Particle Physics at the Tera-scale) and WS and MW by CE200100008 (Dark Matter Particle Physics). ABe is supported by F.N.R.S. through the F.6001.19 convention. ABuc is supported by the Royal Society grant UF160548. JECM is supported by the Carl Trygger Foundation grant no. CTS 17:139. JdB acknowledges support by STFC under grant ST/P001246/1. JE was supported in part by the STFC (UK) and by the Estonian Research Council. BF was supported by EU MSCA-IF project 752162 -- DarkGAMBIT. MF and FK are supported by the  Deutsche Forschungsgemeinschaft (DFG) through the Collaborative Research Center TRR 257 ``Particle Physics Phenomenology after the Higgs Discovery'' under Grant~396021762 -- TRR 257 and FK also under the Emmy Noether Grant No.\ KA 4662/1-1. AF is supported by an NSFC Research Fund for International Young Scientists grant 11950410509. SHe was supported in part by the MEINCOP (Spain) under contract PID2019-110058GB-C21 and in part by the Spanish Agencia Estatal de Investigaci\'on (AEI) through the grant IFT Centro de Excelencia Severo Ochoa SEV-2016-0597. SHoof is supported by the Alexander von Humboldt Foundation. SHoof and MTP are supported by the Federal Ministry of Education and Research of Germany (BMBF). KK is supported in part by the National Science Centre (Poland) under research Grant No. 2017/26/E/ST2/00470, LR under No. 2015/18/A/ST2/00748, and EMS under No. 2017/26/D/ST2/00490. LR and ST are supported by grant AstroCeNT: Particle Astrophysics Science and Technology Centre, carried out within the International Research Agendas programme of the Foundation for Polish Science financed by the European Union under the European Regional Development Fund. MLM acknowledges support from NWO (Netherlands). SM is supported by JSPS KAKENHI Grant Number 17K05429. The work of K.A.O.~was supported in part by DOE grant DE-SC0011842  at the University of Minnesota. JJR is supported by the Swedish Research Council, contract 638-2013-8993. KS was partially supported by the National Science Centre, Poland, under research grants 2017/26/E/ST2/00135 and the Beethoven grants DEC-2016/23/G/ST2/04301. AS is supported by MIUR research grant No. 2017X7X85K and INFN. WS is supported by  KIAS Individual Grant (PG084201) at Korea Institute for Advanced Study. ST is partially supported by the Polish Ministry of Science and Higher Education through its scholarship for young and outstanding scientists (decision no. 1190/E-78/STYP/14/2019). RT was partially supported by STFC under grant number ST/T000791/1. The work of MV is supported by the NSF Grant No.\ PHY-1915005. ACV is supported by the Arthur B. McDonald Canadian Astroparticle Physics Research Institute. Research at Perimeter Institute is supported by the Government of Canada through the Department of Innovation, Science, and Economic Development, and by the Province of Ontario through MEDJCT. LW is supported by the National Natural Science Foundation of China (NNSFC) under grant Nos. 117050934, by Jiangsu Specially Appointed Professor Program.

\section*{Author contributions}

The project was led by AF and in preliminary stages by BF and FK.
ABe, AF, SHoof, AK, PSc and WS contributed to creating the figures.
PA, CB, TB, ABe, ABuc, AF, TEG, SHoof, AK, JECM, MTP, AR, PSc, ACV and YZ contributed to writing.
WH and FK performed official internal reviews of the article.
All authors read, endorsed and discussed the content and recommendations.

\section*{Code availability}
The figures were prepared with \texttt{matplotlib}.\cite{Hunter:2007} We have made all scripts publicly available at Zenodo.\cite{zenodo_record}

\end{document}